\documentclass[letterpaper]{aa}
\usepackage{epsfig,times}
\usepackage{amssymb}
\usepackage{graphicx}

\begin{document}
\title{Magnetorotational instability in proto-neutron stars}
\author{
V.~ Urpin 
}
\institute{
Departament de F\'{\i}sica Aplicada, Universitat d'Alacant,
           Ap. Correus 99, 03080 Alacant, Spain \\
A.F.Ioffe Institute of Physics and Technology, 194021 St.Petersburg, Russia \\
Isaac Newton Institute of Chile, Branch in St.Petersburg, 
194021 St.Petersburg, Russia
}

\date{Received...../ Accepted.....}

\abstract
{Magneto-rotational instability (MRI) has been suggested to lead to a 
rapid growth of the magnetic field in core collapse supernovae and produce 
departures from spherical symmetry that are important in determining the 
explosion mechanism.}
{We address the problem of stability in differentially rotating magnetized
proto-neutron stars at the beginning of their evolution.}
{To do this we consider a linear stability taking into account non-linear 
effects of the magnetic field and strong gravity.}
{Criteria for MRI are derived without simplifying assumptions about a weak 
magnetic field. In proto-neutron stars, these criteria 
differ qualitatively from the standard condition $d \Omega/d s <0$ where
$\Omega$ is the angular velocity and $s$ the cylindrical radius. If the 
magnetic field is strong, the MRI can 
occur only in the neighbourhood of the regions where the spherical radial 
component of the magnetic field vanishes. The growth rate of the MRI
is relatively low except for perturbations with very small scales which usually 
are not detected in numerical simulations. We find that MRI in 
proto-neutron stars grows more slowly than the double diffusive instability 
analogous the Goldreich-Schubert-Fricke instability in ordinary stars.}
{} 




\maketitle

\section{Introduction}

There is a growing amount of evidence that core-collapse supernovae are
asymmetric and that the core-collapse mechanism itself is responsible for 
the asymmetry (see Buras et al. 2003, Akiyama et al. 2003 for more details). 
Several possibilities are explored to account for this observed asymmetry. 
One is associated with the influence of rotation on convection, which seems 
to be inevitable during the early evolution of proto-neutron stars (PNS). 
Convective motions in PNSs are very fast ($\sim 10^{8}-10^{9}$ cm/sec) and,
therefore, the convective turnover time is short, $\sim 1-10$ ms 
(see, e.g., Burrows \& Lattimer 1986). Nevertheless, if angular momentum 
is conserved, the collapsing core can spin up to very short periods $\sim 
5-10$ ms and generate strong differential rotation. Such fast rotation 
modifies convection and makes convective motions anisotropic and constrained 
to the polar regions (Fryer \& Heger 2000, Miralles et al. 2004). This 
mechanism is a natural way to create anisotropic energy and momentum 
transport by convective motions, that only requires that the angular 
velocity be of the order of the Brunt-V\"ais\"al\"a frequency. 

The other possibility to create asymmetry is the effect of jets
(see, e.g., Khokhlov et al. 1999, Wheeler et al. 2002). Even though the 
mechanism of jet formation is still unclear, it seems that MHD jets are 
common in systems where a central body accretes matter with angular momentum 
and magnetic field (see, e.g., Meier at al. 2001), and a core-collapse 
supernova is such a system. Calculations have established that 
nonrelativistic axial jets originating within the collapsed core can 
initiate a bipolar asymmetric supernova explosion that is consistent with 
observations (Hwang et al. 2000). 
 
Another way to generate asymmetry is associated with the magnetic
field that can be an important ingredient of the explosion mechanism
(Bisnovatyi-Kogan 1971, Kundt 1976). The toroidal magnetic field can be 
amplified by differential rotation to such high values that it becomes 
dynamically important (Ardelyan et al. 2005). The effect of the magnetic 
field on asymmetry of supernovae was considered by Wheeler et al.(2000, 
2002) who found that it is possible to produce both a strong toroidal 
field and an axial jet. Two-dimensional MHD simulations of core collapse 
indicate that the shape of shock waves and the neutrinosphere can be modified 
by the effect of the magnetic field (Kotake et al. 2004, Takiwaki et al. 
2004). 

The possible presence of a magnetic field and differential rotation in a
core-collapse supernova favours magnetorotational instability (MRI), 
which can enhance turbulent transport and amplify the magnetic field. 
This instability was considered by Akiyama et al.(2003) in the context of 
core collapse. The authors argued that instability must occur 
in core collapse and that it has the capacity to produce fields that are 
sufficiently strong to affect, if not cause, the explosion. Thompson et al.
(2005) constructed one-dimensional models, including rotation and magnetic
fields, to study the mechanism of energy deposition. They explored several
mechanism for viscosity and argue that turbulent viscosity caused by the MRI
can be most effective. Numerical simulations provide contradictory 
conclusions regarding the importance of MRI in core collapse. Moiseenko 
et al. (2006) claim that MRI has been found in their 2D simulations, and 
that it is responsible for a strong amplification of the poloidal magnetic 
flux. However, what these authors call MRI is different to 
standard MRI considered by Velikhov (1959) (see also Balbus \& Hawley (1991)).
For instance, the instability found by Moiseenko et al (2006) starts to 
develop only when the ratio of the toroidal and poloidal fields reaches a 
value of $\sim$ a few tens. On the other hand, the onset of standard MRI 
in 2D does not depend on the toroidal field at all. The dependence on the 
ratio of the toroidal and poloidal field is more typical for Tayler 
instability (Tayler 1973), which is more relevant to the topology of the 
magnetic field than differential rotation. Therefore, it is possible that 
Moiseenko et al. (2006) incorrectly identify instability, and MRI 
does not occur in their simulations. Apart from that, Moiseenko et al. (2006) 
attributed a rapid growth of the toroidal and poloidal fields to a dynamo 
driven by the magnetorotational instability. This also rise some doubts 
because of Cowlings's anti-dynamo theorem which states that an 
axisymmetric dynamo cannot exist (see, e.g., Shercliff 1965). Two-dimensional 
simulations of core collapse with a strong magnetic field have been 
performed by Sawai et al. (2005, 2008). They found that the magnetic 
field can play an important role in the dynamics of the core only if the 
poloidal field of the progenitor is strong enough ($\sim 10^{12}-10^{13}$ G), 
but MRI was not seen in the considered models. The magnetic field is 
amplified mainly by field compression and field wrapping in these simulations.
On the contrary, Shibata et al. (2006) claim that they found MRI in 
their simulations of magnetorotational core collapse in general relativity. 
These authors paid attention to resolution in order to resolve unstable 
MRI modes and they claim that amplification of the magnetic field in the
considered models is caused by MRI. Note, however, that the poloidal 
field obtained in their models is of the order of that estimated from
conservation of magnetic flux in core collapse, and a more refined 
analysis is required to determine the mechanism of amplification. Fryer \& 
Warren (2004) argued that it is difficult to produce magnetic fields in excess
of $10^{14}$ G even if MRI occurs in core collapse because rotation is not
sufficiently fast. A detailed study of the magnetorotational core collapse has
been performed by Obergaulinger et al. (2006a, 2006b, 2009). The initial 
magnetic 
field was purely poloidal in their models with a strength ranging from 
$10^{10}$ to $10^{13}$ G. Such fields are much higher than those estimated to 
exist in realistic stellar cores, but the authors wanted to investigate the 
principal effects of a magnetic field. The initial magnetic field is
amplified by differential rotation in these simulations, giving rise to a
strong toroidal component. The poloidal component grows mainly by compression
during collapse and does not change significantly after core bounce. The 
authors also found that extended regions exist where the criterion 
of MRI is satisfied at various epochs. However, the growth rate of this 
instability is typically too small, except for a few models
with a strong initial magnetic field $B=10^{12}$ G.

In this paper, we study the effect of MRI on core-collapse supernova. 
Since the magnetic field can be sufficiently strong, the criterion of 
instability is derived taking into account terms depending on the Alfven
frequency. We derive the criterion that applies to any rotation 
profile but a special consideration is made for the case of shellular 
rotation that often is used to mimic rotation of proto-neutron stars.  
We address only the axisymmetric instability because numerical simulations 
of a magnetic core-collapse are usually done in 2D. The main goal of this 
study is to show that the effect of MRI on proto-neutron stars often is 
overestimated.

\section{The growth rate of convective and magnetorotational instabilities}

We assume that the initial PNS is restricted by the radius of neutrino 
sphere. The PNS has a high-entropy mantle, so that the outer part of the star 
is initially at a relatively large radius. It takes a few tenths of a second
for the neutrinos in the high-entropy mantle to leak from the star and for
the mantle to collapse to the canonical radius. In this paper, we study 
MRI in relatively deep layers of the PNS where the density is comparable to
(or higher than) the nuclear density. 

Consider a PNS rotating with angular velocity $\Omega = \Omega(s, z)$;
($s$, $\varphi$, $z$) are cylindrical coordinates. We explore the Boussinesq
approximation and assume that the magnetic energy is small compared to the 
thermal one. In the unperturbed state, the star is in hydrostatic equilibrium,
\begin{equation}
\frac{\nabla p}{\rho} = \vec{G} + \frac{1}{4 \pi \rho} (\nabla 
\times \vec{B}) \times \vec{B} \; , \;\;\;
\vec{G} = \vec{g} + \Omega^{2} \vec{s} 
 \end{equation}
where $\vec{g}$ is the gravity. Generally, the magnetic field $\vec{B}$ has 
both toroidal $\vec{B}_{\varphi}$ and poloidal $\vec{B}_{p}$ components. 
If the field has a component parallel to $\nabla \Omega$, the azimuthal field 
in the unperturbed state increases with time by winding up the polodal field 
lines. If the magnetic Reynolds number is large, then
$$
B_{\varphi}(t) = B_{\varphi}(0) + s (\vec{B}_{p} \cdot \nabla \Omega) t .
$$  
We consider MRI assuming that the basic state is quasi-stationary. This 
is justified if the growth rate of the MRI ($\sim \Omega$) is greater than 
the inverse time-scale on which the basic state evolves. This inverse 
time-scale can be estimated as $|\dot{p}/p| \sim s |\nabla \Omega| 
(p_{M}/p) \sim (p_{M}/p) \Omega $, where $p_{M}$ is the magnetic pressure. 
Therefore, the condition of quasi-stationarity is satisfied if $p_{M}/p 
\leq 1$, that is fulfilled in core collapse supernovae. 

We consider axisymmetric short-wavelength 
perturbations with spatial and temporal dependence $\exp(\gamma t - i \vec{k} 
\cdot \vec{r})$ where $\vec{k}= (k_{s}, 0, k_{z})$ is the wave-vector. Small 
perturbations will be indicated by a subscript 1. Then, the linearized MHD 
equations read in a short-wavelength approximation
\begin{eqnarray}
\gamma \vec{v}_{1} + 2 \vec{\Omega} \times \vec{v}_{1} +
\vec{e}_{\varphi} s (\vec{v}_{1} \nabla \Omega) =
\frac{i \vec{k} p_{1}}{\rho} + \vec{G} \; \frac{\rho_{1}}{\rho} + 
\nonumber \\
\frac{i}{4 \pi \rho} [\vec{k} (\vec{B} \cdot \vec{B}_{1}) -
\vec{B}_{1} (\vec{k} \cdot \vec{B})] \;,
\end{eqnarray}
\begin{equation}
\vec{k} \cdot \vec{v}_{1} = 0 \;, 
\end{equation}
\begin{equation}
\gamma \vec{B}_{1} = - i \vec{v}_{1} (\vec{k} \cdot \vec{B})
+ s \vec{e}_{\varphi} (\vec{B}_{1} \cdot \nabla \Omega) \;, 
\end{equation}
\begin{equation}
\vec{k} \cdot \vec{B}_{1} = 0 \;,
\end{equation}
where $\vec{e}_{\varphi}$ is the unit vector in the $\varphi$-direction. 
We neglect kinematic and magnetic viscosities since they are small in PNSs.
The term $\propto \nabla \times \vec{B}$ in Eq.~(2)) is also neglected. In the 
short-wavelength approximation, the latter is justified if $B_{\varphi}$ 
satisfies the inequality $B_{\varphi} < (L/\lambda) B_{p}$, where $L$ is the 
length scale of unperturbed quantities and $\lambda = 2 \pi/ k$. 

The matter is assumed to be in chemical equilibrium, thus the density 
is a function of the pressure $p$, temperature $T$ and lepton fraction $Y$. 
In the Boussinesq approximation, perturbations of the pressure are small and, 
therefore, $\rho_1$ can be expressed in terms of $T_1$ and $Y_1$,
\begin{equation}
\rho_{1} \approx - \rho \beta (T_{1}/T) - \rho \delta Y_{1},
\end{equation} 
where $\beta$ and $\delta$ are the coefficients of thermal and chemical
expansion; $\beta = - (\partial \ln \rho/\partial \ln T)_{pY}$, $\delta =
- (\partial \ln \rho/ \partial Y)_{pT}$.
Since diffusive timescales are long compared to dynamical ones (Miralles 
et al. 2000), the linearized transport equations read
\begin{equation}
{\dot{T}_{1}} - \vec{v}_{1} \cdot  {\Delta\nabla T} = 0,
\label{temp} 
\end{equation}
\begin{equation}
\dot{Y}_{1} + \vec{v}_{1} \cdot \nabla Y = 0,
\label{lept} 
\end{equation}
where $\Delta \nabla T= ( \partial T /\partial p)_{s, Y}
\nabla p - \nabla T$ is the super-adiabatic temperature gradient.

The dispersion equation corresponding to Eqs. (2)--(8) is
\begin{equation}
\gamma^{4} + b_{2} \gamma^{2} + b_{0} = 0,
\end{equation}
where
\begin{eqnarray}
b_{2} = 2 \omega_{A}^{2} + \omega_{c}^{2} + q^{2} ,\;\;\;
b_{0} = \omega_{A}^{2} ( \omega_{A}^{2} + \omega_{c}^{2} + q^{2}
- 4 \Omega^{2} k_{z}^{2}/k^{2} ) .
\nonumber
\end{eqnarray}
Here $\omega_{A} = (\vec{k} \cdot \vec{B})/\sqrt{4 \pi \rho}$ is the Alfven
frequency and
$$
q^{2} = (k_{z}^{2} \Omega_{e}^{2} - s k_{s} k_{z}\Omega^{2}_z)/k^{2} ,\;\;\; 
\omega_{c}^{2} = - \vec{C} \cdot [ \vec{G} - \vec{k}
(\vec{k} \cdot \vec{G})/k^{2} ] ,   
$$
where $\vec{C} = - (\beta/T) \Delta\nabla T + \delta \nabla Y$, $\Omega_{e}$ 
is the epicyclic frequency, $\Omega_{e}^{2} = \partial (s^{4} \Omega^{2})/ 
s^{3} \partial s$, and $\Omega_{z}^{2} =  \partial \Omega^{2}/\partial z$. 

Eq.~(9) has an unstable solution if either $b_{2}$ or $b_{0}$ is negative, or 
\begin{equation}
2 \omega_{A}^{2} + \omega_{c}^{2} + q^{2} < 0, \;\;\;\;
\omega_{A}^{2} + \omega_{c}^{2} + q^{2} - 4 \Omega^{2} (k_{z}^{2}/k^{2}) 
< 0 . 
\end{equation}
In the limit of a weak field, these conditions yield
\begin{equation}
\omega_{c}^{2} + q^{2} < 0, \;\;\;\;\;\;\;
\omega_{c}^{2} + q^{2} - 4 \Omega^{2} (k_{z}^{2}/k^{2}) 
< 0 . 
\end{equation}
The first inequality is the criterion of convection modified by
rotation, and the second condition represents the criterion of MRI.  

The solution of Eq.~(9) is
\begin{equation}
\gamma^{2} \! = \! -\omega_{A}^{2} - \frac{1}{2} (\omega_{c}^{2}+ q^{2})
\pm \left[ \frac{1}{4} (\omega_{c}^{2} + q^{2})^{2} \! + \! 4 \omega_{A}^{2}
\Omega^{2} \frac{k_{z}^{2}}{k^{2}} \right]^{1/2} \!.
\end{equation}

If $\omega_{A}$ is small compared to the other frequencies, we have
\begin{equation}
\gamma_{1,2}^{2} = - (\omega_{c}^{2} + q^{2}) , \;\;\;\;\;\;
\gamma_{3,4}^{2} = - \omega_{A}^{2} \left[ 1 - \frac{4 \Omega^2 k_z^2}{k^2 
(\omega_{c}^2 + q^2)} \right] .
\end{equation}
The solutions $\gamma_{1,2}$ correspond to buoyant modes that cause 
convection and are unstable if $\omega_{c}^{2} + q^{2} <0$. The solutions 
$\gamma_{3,4}$ describe the magnetorotational modes which can be unstable if 
the second condition (12) is satisfied. 

For the purpose of illustration, we plot in Fig.~1 the dependence of 
$\gamma^2$
on $\omega_{A}/\Omega$ for the unstable magnetorotational mode. Its growth 
rate is given by Eq.(12) with the upper sign. Even if stratification is 
negligible, the growth rate of this mode is typically low and $\gamma \sim 
\omega_{A}$ in a weak magnetic field ($\omega_{A} < \Omega$). Stratification 
can substantially decrease the growth rate and this is seen very well from
the figure. For example, the growth rate decreases approximately by a factor 
of two if $\omega_{c}^2/\Omega^2$. At $\omega_{c}^2/ \Omega^2 \approx 1$,
stable stratification completely suppresses the magnetorotational instability 
of the considered perturbations.     

\begin{figure}
\includegraphics[width=6.5cm, angle=270]{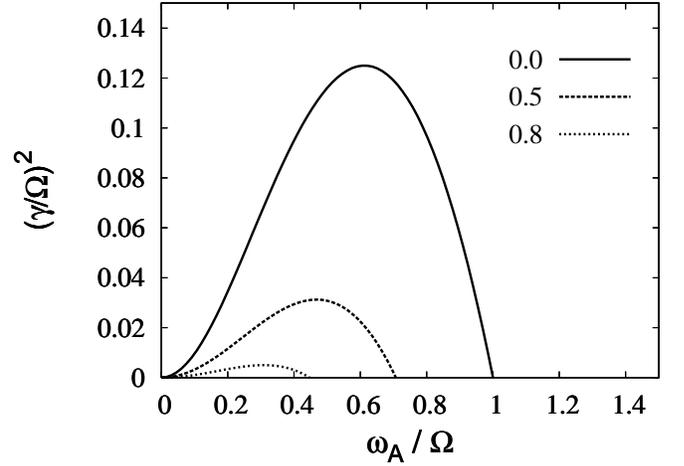}
\caption{The dependence of $\gamma^2$ on $\omega_{A}/ \Omega$ for $\Omega
\propto 1/s$ and $k_z/k=\sqrt{2}/2$ and for three values of the ratio
$\omega_{c}^2/ \Omega^2 = 0.0$, $0.5$, and $0.8$.}
\end{figure}

In the convective zone where $\omega_c^2 + q^2 < 0$, the quantity 
$\omega_c^2 + q^2 - 4 \Omega^2 (k_z^2/k^2)$ is also 
negative and, hence, $\gamma^2_{3,4}< 0$. Therefore, the magnetorotational 
instability does not occur in convectively unstable regions, and these two
instabilities are spatially separated if the magnetic field is weak (see also
Obergaullinger et al. 2009).       
The value of $\omega_{c}$ is of the order of 1-10 ms$^{-1}$ 
in collapsing cores (see Thompson et al. 2005). Therefore, rotation has an 
important impact on convection only if $\Omega$ is of 
the same order of magnitude, $\Omega \approx 1000$ rad/s (Miralles et al. 
2004). This value can be reached in PNS if 
rotation of the progenitor was very fast (Villain et al. 2004). On the 
contrary, MRI can have an important influence even 
if rotation is slower.

\section{The criteria of magnetorotational instability}

The condition of MRI (second inequality (10)) 
depends on the direction of $\vec{k}$ and can be written as follows
\begin{equation}
F \equiv \frac{b_{0}}{\omega_{A}^{2}} = \omega_{0}^{2} + 
A \frac{k_{z}^{2}}{k^{2}} - D \frac{k_{s} k_{z}}{k^{2}} + 
E \frac{k_{s}^{2}}{k^{2}} < 0,
\end{equation}
where
\begin{eqnarray}
&& A = s \Omega_{s}^{2} + \cos^{2} \alpha \omega_{A0}^{2} + C_{z} G_{z}, 
\;\;\; E = \sin^{2} \alpha \omega_{A0}^{2} + C_{s} G_{z},  
\nonumber \\
&& D = s \Omega_{z}^{2} - \sin 2\alpha \ \omega_{A0}^{2} -C_{s} G_{z}
- C_{z} G_{s} . \nonumber
\end{eqnarray}
In these expressions, we denote
\begin{eqnarray}
\omega_{0}^{2} = - \vec{C} \cdot \vec{G} \;, \;\; 
\omega_{A0}^{2} = \frac{k^{2} B^{2}}{4 \pi \rho} \;, \;\;
\Omega_{s}^{2} = \frac{\partial \Omega^{2}}{\partial s} \; ;
\end{eqnarray}
$\alpha$ is the angle between the magnetic field and the rotational axis,
$\cos \alpha = B_z / \sqrt{B_z^2 + B_s^2}$.

Since the dependence of $F$ on the direction of $\vec{k}$ is
simple, we can obtain that $F$ reaches its minimum at 
\begin{equation}
\frac{k_{z}^{2}}{k^{2}} = \frac{1}{2} 
\left[ 1 \pm \sqrt{\frac{(A-E)^{2}}{(A-E)^2 + D^2}} \right]~.
\label{phimax}
\end{equation}
The value of $F$ corresponding to these $k_{z}^{2}/k^{2}$
yields the following condition of instability
\begin{equation}
s \Omega_{s}^{2} + \omega_{0}^{2} + \omega_{A0}^{2} \pm
\sqrt{ D^{2} + (A-E)^{2}} < 0. 
\end{equation}
By taking the curl of Eq. (1), it can be readily obtained that 
the condition of hydrostatic equilibrium leads to 
\begin{equation}
s \Omega_{z}^{2} = \left[\vec{C}\times\vec{G}_{B} - \vec{L} \right]_{\varphi} =
C_z G_{Bs} -C_s G_{Bz} -L_{\varphi},
\label{eqh2}
\end{equation}
where
$$
\vec{G}_{B} = \vec{G} + \frac{1}{4 \pi \rho} (\nabla \times \vec{B}) \times 
\vec{B} \;,  \;\;\; \vec{L} = \nabla \times (\vec{G}_{B} - \vec{G}).
$$
We consider only the magnetic field satisfying the condition $L \ll s 
\Omega_s^2$ that is approximately equivalent to the requirement that the 
magnetic energy is small compared to the rotational energy. Then, Eq. (17) 
can be further simplified to obtain
\begin{eqnarray}
s \Omega_{s}^{2} + \omega_{0}^{2} + \omega_{A0}^{2} \pm
\left\{ (s \Omega_{s}^{2} + \omega_{0}^{2} + \omega_{A0}^{2})^{2} + 
\right.
\nonumber \\
G_{z} [ s(C_{z} \Omega_{s}^{2} - C_{s} \Omega_{z}^{2}) +
\omega_{A0}^{2} ( C_{z} \cos 2 \alpha  + C_{s} \sin 2 \alpha)] - 
\nonumber \\
\left.
\sin^{2} \alpha \omega_{A0}^{2} (\omega_{0}^{2} + s \Omega_{s}^{2})
\right\} < 0. 
\end{eqnarray} 
The two conditions of instability follow straightforwardly from the above 
expression:
\begin{eqnarray}
s \Omega_{s}^{2} + \omega_0^2 + \omega_{A0}^{2} < 0~, \\
s G_z (C_z \Omega_{s}^{2} - C_s \Omega_{z}^{2}) + 
\omega_{A0}^{2} [ C_z G_z \cos^{2} \alpha 
+ C_s G_z \sin 2 \alpha  
\nonumber \\
+ (C_s G_s - s \Omega_{s}^{2}) \sin^{2} \alpha] > 0~.
\label{cond2} 
\end{eqnarray}
Conditions (20) and (21) look like the Solberg--H{\o}iland conditions
(Tassoul 2000), but with additional terms due to the chemical
composition gradients and the magnetic field. If the magnetic field is
weak and $\nabla Y=0$ then Eqs. (20) and (21) yield
\begin{equation}
s \Omega_{s}^{2} + \omega_0^2  < 0~; \\
s G_z (C_z \Omega_{s}^{2} - C_s \Omega_{z}^{2}) > 0~. 
\end{equation}  
The first criterion is similar to the Schwarzschild criterion for convection 
modified by rotation. However, the Schwarzschild criterion at $B=0$ 
involves the angular momentum gradient, whereas the criterion of  
magnetorotational instability depends on the angular velocity 
gradient, as was noted by Balbus (1995). If rotation is cylindrical 
with $\Omega^{2}_{z}= 0$, the second criterion (22) yields the standard 
condition of the MRI, $\Omega^{2}_{s} < 0$, since $C_z G_z < 0$ in a 
convectively stable region. 

Taking into account that $C_z \Omega_s^2 - C_s \Omega_z^2 = (\vec{C} \times 
\nabla \Omega^{2})_{\varphi}$, Eq.(21) can be transformed into 
\begin{eqnarray}
s G_z |C| |\nabla \Omega^2| \sin \psi + 
\omega_{A0}^{2} [ C_z G_z \cos^{2} \alpha 
+ C_s G_z \sin 2 \alpha  
\nonumber \\
+ (C_s G_s - s \Omega_{s}^{2}) \sin^{2} \alpha] > 0~,
\end{eqnarray}
where $\psi$ is the angle between vectors $\vec{C}$ and $\nabla \Omega^2$.
If the Alfven frequency is small, then the criterion of MRI reads
\begin{equation}
0 > \psi > - \pi,
\end{equation}
and is different from the usually used condition $\Omega^2_s < 0$. Criterion
(24) implies that the component of $\nabla \Omega^2$ perpendicular to 
$\vec{C}$ must have a negative projection on $\vec{e}_s$ for instability
whereas the component along $\vec{C}$ plays no role.  In the general case 
when $\omega_{A0}$ is comparable to (or greater than) other frequencies 
(generally, that is possible for very 
short wavelengths even if $B$ is weak), criterion 
(23) can be complicated. For instance, if  
$\omega_{A0} \gg \Omega$, then criterion (23) yields
\begin{equation} 
C_z G_z \cos^{2} \alpha + C_s G_z \sin 2 \alpha  
+ (C_s G_s - s \Omega_{s}^{2}) \sin^{2} \alpha > 0~.
\end{equation}
This condition depends on the direction of $\vec{B}$ and is different 
from the standard condition $\Omega_s < 0$. Note, for example, that this 
condition cannot be satisfied if  
$B_s=0$ because in this case $\sin \alpha =0$ and $C_z G_z < 0$ in a 
convectively stable region.

\section{Instability in core-collapse supernovae}

The occurence of MRI is sensitive to the rotation profile. It follows
from both theoretical modelling and analytic consideration that core 
collapse of a rotating progenitor results in differential rotation of a 
protoneutron star (Zwerger \& M\"uller 1997, Dimmelmeier et al. 2002, 
M\"uller et al. 2004, Obergaulinger et al. 2006a,b). Many studies of MRI 
model rotation of the 
collapsing core by a shellular profile with $\Omega = \Omega (r)$ where $r$ 
is the spherical radius (see, e.g., Akiyama et al. 2003, Thompson et al.2005, 
Sawai et al. 2005). Such rotation can be justified if the progenitor 
rotates with an angular velocity that depends on $r$ alone (M\"onchmeyer 
\& M\"uller 1989). Then, if the angular momentum is conserved, it can be 
shown that rotation of the collapsing core is shellular at the beginning of 
evolution, at least (see, e.g., Akiyama et al. 2003).  However, there are
studies that assume a cylindrical rotation for the initial profile (see, e.g.,
Obergaulinger et al. 2006a,b). 
 
We assume that the angular velocity is low compared to the Keplerian
one and little departure occurs from a spherical geometry. Then, $\vec{C}$ 
and $\vec{G}$ are approximately radial, and we have
\begin{equation}
G_s \!=\! -\! g \sin \theta , \; G_z \! =\! - \! g \cos \theta, \; 
C_s \!=\! C \sin \theta , \; C_z \!=\! C \cos \theta, 
\end{equation}
where $\theta$ is the polar angle. The quantity $C$
is negative in a convectively unstable region and positive in a
stable region. 

Substituting expressions (26) into Eqs. (20)-(21), we obtain
\begin{eqnarray}
s \Omega_{s}^{2} + \omega_0^2 + \omega_{A0}^{2} < 0~; \\
s \omega_{0}^{2} \cos \theta ( \cos \theta \Omega_{s}^{2} - \sin \theta
\Omega_{z}^{2} ) +  \nonumber  \\ 
\omega_{A0}^{2} [ \omega_{0}^{2} (\cos \theta
\cos \alpha + \sin \theta \sin \alpha )^{2} + s \Omega_{s}^{2} 
\sin^{2} \alpha ] < 0.
\end{eqnarray}
Condition (28) can be satisfied only if the angular velocity is higher 
than the Brunt-V\"ais\"al\"a frequency which is rather high in core-collapse
supernovae. For example, according to calculations by Thompson et al. (2005), 
the value of $\omega_0$ decreases from $\sim 10^3$ to $10^2$ s$^{-1}$ as the 
radius increases from 50 to 200 km in the gain region.
The angular velocity can be comparable only if the progenitor rotated
very rapidly with a period $\sim 2$ s. Such rotation can be 
achieved only in the most rapidly rotating stars, and our study does not 
address
such stars. We concentrate on condition (28) that can be satisfied for
much slower rotation. 

If rotation is approximately shellular as is often assumed and $\Omega 
\approx \Omega(r)$, then we have $\Omega^2_s = \Omega_r^2 \sin \theta$ and 
$\Omega^2_z = \Omega_r^2 \cos \theta$ where $\Omega^2_r = \partial \Omega^2/ 
\partial r$. Then, Eq.(28) can be transformed into
\begin{equation}
\omega^2_0 \cos^2 (\theta -\alpha) + r \sin^2 \theta \sin^2 \alpha
\; \Omega^2_r < 0 .
\end{equation}  
It turns out that stratification strongly suppresses MRI for such a 
rotation profile because $\omega^2_0 > 0$ in convectively stable regions 
and, typically, $\omega^2_0 > \Omega^2$. 

\begin{figure}
\includegraphics[width=6.5cm, angle=270]{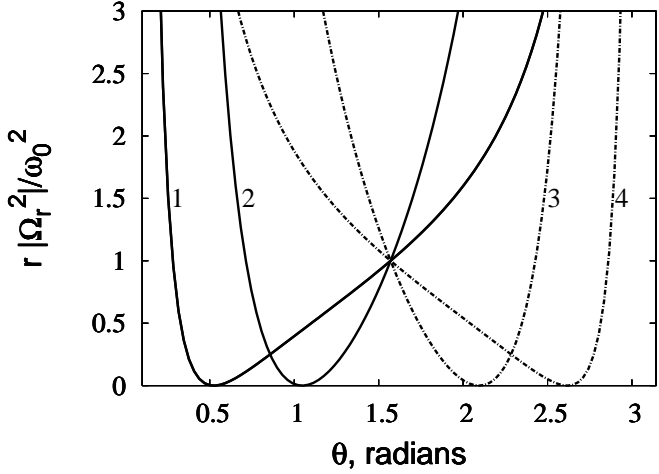}
\caption{The dependence of the critical value of $r |\Omega_r^2|$ that 
determines the instability on the polar angle for $\alpha = - \pi/3$ 
(curve 1). $-\pi/6$ (2), $\pi/6$ (3), and $\pi/3$ (4). 
The regions above the lines correspond to instability.}
\end{figure}

In Fig.~2, we plot the critical value of $r\Omega_r^2$ that discriminates 
between stable and unstable regions as a function of $\theta$ for several 
values of the angle $\alpha$. The region above the corresponding curve is 
magnetorotationally unstable. 
For any given $\alpha$ and negative $\Omega_r^2$, there exists a range of
$\theta$ where the instability can arise. If $r \Omega_r^2$ is greater than
(or comparable to) $\omega_0^2$ then the instability occurs over a rather wide
range of $\theta$. If $r |\Omega_r^2| \ll \omega_0^2$ (which is more typical 
for PNSs) then the instability arises only in a very narrow range of $\theta$.
This dependence can be easily understood from Eq.~(29). Even if
$\Omega^2_r < 0$, MRI occurs only in those regions where the first term 
on the l.h.s. of Eq.~(29) is small. This occurs in the neighbourhood of the 
line 
\begin{equation}
\theta = \alpha \pm \pi / 2 .
\end{equation} 
Condition (30) implies that the field line is perpendicular to the 
radius and, hence, at any magnetic topology, the instability occurs
only near the region where $B_r = 0$. How extended this region is
depends on the relation between $\omega^2_0$ and $r \Omega^2_r$.     

There is no generally accepted point of view regarding topology of the 
magnetic field in core-collapse supernovae. For illustration, 
we consider the simplest configuration with
\begin{equation}
B_{r} = f(r) (1 - 3 \cos^{2} \theta), \;\;\; B_{\theta} =
F(r) \sin \theta \cos \theta,
\end{equation}
where $f$ and $F$ are functions of the spherical radius that satisfy 
the divergence-free condition $F = d(r^{2} f)/r dr$.
This field can be matched to the dipole  component outside the
star (see, e.g., Urpin et al. 1994). Generally, the field can have a more 
complex geometry but a simple model (31) allows us to show qualitatively 
how the magnetic topology infuences the region of MRI.
We have for the magnetic field (31) 
\begin{eqnarray}
\sin \alpha = \frac{\vec{e}_s \cdot \vec{B}}{B_{p}} =
\frac{\sin \theta}{B}[f (1 - 3 \cos^{2} \theta) + F \cos^{2}
\theta ],     \nonumber \\
\cos \alpha = \frac{\vec{e}_z \cdot \vec{B}}{B_{p}} =
\frac{\cos \theta}{B}[f (1 - 3 \cos^{2} \theta) - F \sin^{2}
\theta ].
\end{eqnarray}
Substituting these expressions into Eq. (29), we obtain 
\begin{equation}
\omega_{0}^2 (1 - 3 \cos^{2} \theta)^2 + r \Omega_{r}^{2} \sin^4 \theta
[ 1 - (1 - x) \cos^{2} \theta]^2 < 0,
\end{equation}
where $x = d \ln f / d \ln r$. Obviously, this inequality can be satisfied
only if the angular velocity decreases with $r$. If $\Omega_r^2 < 0$, then
criterion (33) is always fulfilled at $\cos \theta \approx 1/\sqrt{3}$ (or 
at $\theta \approx \theta_0 =54^{\circ}$) except for the case $x =-2$. Hence, 
some region in the neighbourhood of this polar angle can also be unstable, 
and the opening angle of the unstable region depends on the ratio 
$|r \Omega_r^2/ \omega_0^{2}|$. For example, if this ratio is small then 
MRI occurs within the cone $\theta_0 + \Delta \theta > \theta > 
\theta_0 - \Delta \theta$, where  
\begin{equation}
\Delta \theta \approx \frac{|2+x|}{3\sqrt{2}} 
\frac{\sqrt{|r \Omega_r^2|}}{\omega_0}.
\end{equation}

\section{Conclusion}

Interest in MRI  in core collapse 
supernova is due to the fact that it can generate a strong 
magnetic field and produce significant departures from spherical 
symmetry. These departures are crucial for the explosion mechanism. In 
the present paper, we have derived criteria of MRI in proto-neutron 
stars taking into account the effect of a non-axial magnetic field. 
Criteria have been obtained in a form analogous to the Solberg--H{\o}iland 
criteria but including terms containing the magnetic field. It turns
out that the criterion of MRI in proto-neutron stars can differ
from the standard condition $\partial \Omega / \partial s < 0$
even in a weak magnetic field due to strong gravity and the gradient of the
lepton fraction. If the Alfven frequency $\omega_{A0}$ is 
small compared to the angular velocity then the instability occurs in 
the region where the component of $\nabla \Omega^2$ perpendicular to 
$\vec{C}= -(\beta/T) \Delta \nabla T + \delta \nabla Y$ has a negative 
projection on $\vec{e}_s$ whereas the component $\nabla \Omega^2$ parallel 
to $\vec{C}$ is unimportant for instability. In the case of slow rotation, 
when departures from sphericity are small ($g \gg s \Omega^2$), the criterion 
reduces to a simple inequality $\partial \Omega/ \partial \theta < 0$ 
(see, e.g., Urpin 1996). For instance, shellular rotation with $\Omega=
\Omega(r)$ which is often used in modelling proto-neutron stars does not 
satisfy this condition. Therefore, MRI does not occur if the 
proto-neutron star rotates shellularly and the wavelength of perturbations 
is such that the Alfven frequency is smaller than $\Omega$. Only detailed 
calculations of rotational core collapse can give the answer to whether the 
condition of MRI is fulfilled in proto-neutron stars and, generally, 
this answer should depend on rotation of the progenitor. Note that the 
velocity of unstable perturbations is approximatelly perpendicular to the 
radius because gravity strongly suppresses motion in the radial direction
and, likely, the radial turbulent transport should be suppressed when the 
instability saturates. 

However, MRI can arise in proto-neutron stars even if the neccesary
condition that determines the onset of instability in a weak field is not 
fullfied. This occurs if the magnetic field is very strong or the
wavelength of perturbations is small such that $\omega_{A0} > \Omega$.
The latter inquality is equivalent to
\begin{equation}
\lambda < \lambda_c =2 \pi B/ \Omega \sqrt{4 \pi \rho} \approx 1.8 \times 10^3 B_{13}
\Omega_3^{-1} \rho_{14}^{-1/2} \;\; {\rm cm}, 
\end{equation}
where $\lambda =2\pi/k$ is the wavelength, $B_{13}=B/10^{13}$G, $\Omega_3 =
\Omega/10^3$ s$^{-1}$, $\rho_{14}= \rho/10^{14}$ g/cm$^3$. For perturbations,
satisfying condition (35), the instability can occur if $\Omega_s < 0$ but
only in the neighbourhood of the cone (or cones depending on the field 
topology) 
in which the radial component of $\vec{B}$ is vanishing. The opening angle 
of the unstable region around this cone is of the order of $\Omega/\omega_0$ 
(see Eq.(34)). Note that unstable motion in this case are perpendicular to 
the radius as well since gravity suppresses radial motions. The reason why 
the instability is not entirely suppressed even by a very strong magnetic 
field is qualitatively very simple. Perturbations with $\vec{k}=(k_r, 0, 0)$ 
cannot be suppressed by gravity, and such perturbations do not feel the 
stabilizing influence of the magnetic field in the region where $B_r \approx 
0$ as well, because this influence  is proportional to the Alfven frequency 
($\propto \vec{k} \cdot \vec{B}$). 

It appears that the importance of MRI in core collapse can be 
overappreciated since its growth rate is relatively low and reaches the value 
$\sim \Omega$ only for perturbations with a wavelength $\sim \lambda_c$ 
(see Eq.(35)). Apart from this, gravity strongly suppresses development of 
any perturbations with non-radial wavevectors. Only perturbations with 
$k \approx k_r$ can be unstable but hydrodynamic motion for such 
perturbations has a small radial component and turbulent transport should 
be inefficient radially. Also, it is possible that  MRI occurs only in 
not very extended regions of the proto-neutron star that may diminish 
substantially its effect on core collapse. Note that since the wavelength 
$\lambda_c$ is small, one needs a very high resolution in numerical simulations
to see the most rapidly growing modes (Obergaulinger et al. 2009). 
Perturbations with longer $\lambda$ grow substantially slower. 

Our conclusion is in contrast to the results obtained by Masada et al. 
(2007) who considered axisymmetric and nonaxisymmetric magnetorotational 
instability of PNSs taking into account dissipative processes. These authors 
also used a local approximation in the stability analysis. In the local 
analysis, however, the shape of perturbations is assumed to be unchanged and, 
therefore, this approach applies only if the growth rate of instability is
greater than the rate with which perturbations change their shape. In the
case of MRI, the growth rate is smaller than (or, at maximum, comparable 
to) $s |\nabla \Omega|$ whereas perturbations change their shape because of 
differential rotation with a rate $\sim s |\nabla \Omega|$ (see, e.g., 
Balbus \& Hawley 1992 for details). Therefore, the results obtained for a
nonaxisymmetric MRI in the local approximation raise some doubts. As far as
axisymmetric instability is concerned, Masada et al. (2007) were confused 
when identifying different modes in the dispersion equation. In the 
dissipative case, they obtained a dispersion equation of the seventh order 
which describes seven different modes. Certainly, different modes can be 
unstable 
in different conditions, but the authors considered only one criterion 
(Eq.(32) of their paper). Unfortunately, this criterion does not correspond 
to MRI but describes a secular instability that is a magnetic analogy of 
the well known Goldreich-Schubert-Fricke (GSF) instability. The magnetic 
analogy of this instability was first considered by Urpin (2006) for the case 
of ordinary stars, and Masada et al. (2007) obtained the criterion of the
same instability modified for the conditions of PNSs. In contrast to MRI, 
this instability is dissipative and disappears if diffusive coefficients go
to zero. Stratification has a weak impact on this dissipative instability, 
and it can arise in PNSs. Note, however, that diffisive processes can reduce
the stabilizing influence of stratification on the MRI as well.

As was
noted by Masada et al. (2007), diffusion of heat and leptons can influence 
buoyancy if the characteristic diffusion timescale is shorter than the
buoyant frequency $\omega_0$. Since the diffusion timescale is determined 
mainly by lepton diffusion which is slower than thermal diffusion, this
condition is equivalent to $\xi k^2 \geq \omega_0$ where $\xi$ is the 
coefficient of lepton diffusivity. In deep layers with a density 
$\sim 10^{14}$ g/cm$^3$, the value of $\xi$ is $\sim 3 \times 10^7$ 
cm$^2$s$^{-1}$ (see Eq.(45) by Masada et al. (2007)). Hence, diffusion 
begins to suppress the stabilizing effect of stratification when the 
wavelength of perturbations is shorter than $\sim 2 \pi \sqrt{\xi / \omega_0} 
\sim 10^3$ cm (we suppose $\omega_0 \sim 10^3$ s$^{-1}$). The effects 
considered in this paper are important for longer wavelengths for the standard 
pulsar magnetic field.

As was mentioned above, MRI is not the only instability caused by 
differential rotation in proto-neutron stars. The analogy of the GSF 
dissipative instability can occur in both magnetic and non-magnetic PNSs.
In non-magnetic stars, this instability arises if the angular velocity depends
on the vertical coordinate $z$ (Urpin 2007), and the shellular rotation is a 
particular case of such rotation. Since neutrino transport in PNSs is much 
more efficient than radiative transport in ordinary stars, the dissipative
instability can be rather fast. If $\Omega_z= \partial \Omega / \partial z 
\sim \Omega/ s$, the condition of the instability reads 
\begin{equation}
\frac{|s \Omega_z|}{\sqrt{\Omega_e^2}} \sim \Omega \gg 2 \omega_0 
\sqrt{\frac{\nu}{\kappa_T}},
\end{equation}
where $\nu$ and $\kappa_T$ are the coefficients of viscosity and thermal
diffusivity. The difference to the standard 
criterion of GSF instability ($\partial \Omega/ \partial z \neq 0$)
is caused by the more complex character of neutrino trancport that involves
diffusion of both heat and the lepton fraction. This instability can arise 
soon after collapse even in dense central regions of the star where 
convection does not occur. It turns out that in all cases, when  
dissipative instability arises, its growth rate is higher than that of  
MRI. Indeed, the maximum growth rate of MRI is $\sim \Omega$ and can be 
reached for perturbations with  wavelength $\sim \lambda_c$. The
growth rate of the dissipative instability is approximately $\kappa_T k^2
(\Omega / \omega_0)^2$. The maximum wavevector for which the instability 
occurs is $\sim \sqrt{\Omega/\nu}$. For perturbations with such wavevector,
the growth rate is of the order of $\Omega (\kappa_T /\nu)(\Omega/\omega_0)^2$
and is always higher than the growth rate of  MRI if condition (36) is
satisfied. Therefore, dissipative instability seems to be  more efficient 
in proto-neutron stars than MRI.

\end{document}